\documentclass[11pt]{article}
\usepackage{amsmath}
\usepackage{xspace}
\usepackage{amssymb}
\usepackage{epsfig}

\newtheorem{Thm}{Theorem}

\newtheorem{Cor}[Thm]{Corollary}
\newtheorem{Prop}[Thm]{Proposition}

\newtheorem{Def}{Definition}
\newenvironment{proof}{\noindent {\textbf{Proof }}}{$\Box$ \medskip}

\newcommand\B{\{0,1\}}

\newcommand\ket[1]{| #1 \rangle}

\newcommand\qip[2]{\langle #1 | #2 \rangle}

\newcommand {\etal} {\emph{et al}\xspace}
\newcommand {\ie} {\emph{i.e.}\xspace}
\newcommand {\st} {\emph{s.t.}\xspace}

\begin{document}
\title{On the power of Ambainis's lower bounds \thanks{This research
was supported in part by NSF grant CCR-0310466. A preliminary
report of the present paper is at arXiv:quant-ph/0311060}}

\author{Shengyu Zhang \\ Computer Science Department, Princeton
University, NJ 08544, USA. \\ Email: szhang@cs.princeton.edu}
\date{}
\maketitle

\pagenumbering{arabic} \pagestyle{plain}

\abstract{The polynomial method and the Ambainis's lower bound (or
\emph{Alb}, for short) method are two main quantum lower bound
techniques. While recently Ambainis showed that the polynomial
method is not tight, the present paper aims at studying the power
and limitation of \emph{Alb}'s. We first use known \emph{Alb}'s to
derive $\Omega(n^{1.5})$ lower bounds for \textsc{Bipartiteness},
\textsc{Bipartiteness Matching} and \textsc{Graph Matching}, in
which the lower bound for \textsc{Bipartiteness} improves the
previous $\Omega(n)$ one. We then show that all the three known
Ambainis's lower bounds have a limitation $\sqrt{N\cdot
\min\{C_0(f), C_1(f)\}}$, where $C_0(f)$ and $C_1(f)$ are the 0-
and 1-certificate complexity, respectively. This implies that for
some problems such as \textsc{Triangle}, $k$-\textsc{Clique}, and
\textsc{Bipartite/Graph Matching} which draw wide interest and
whose quantum query complexities are still open, the best known
lower bounds cannot be further improved by using Ambainis's
techniques. Another consequence is that all the Ambainis's lower
bounds are not tight. For total functions, this upper bound for
\emph{Alb}'s can be further improved to $\min
\{\sqrt{C_0(f)C_1(f)}, \sqrt{N\cdot CI(f)}\}$, where $CI(f)$ is
the size of max intersection of a 0-and a 1-certificate set. Again
this implies that $Alb$'s cannot improve the best known lower
bound for some specific problems such as \textsc{And-Or Tree},
whose precise quantum query complexity is still open. Finally, we
generalize the three known \emph{Alb}'s and give a new \emph{Alb}
style lower bound method, which may be easier to use for some
problems.
\section{Introduction}
Quantum computing has received a great deal of attention in the
last decade because of the potentially high speedup over the
classical computation. Among others, query model is extensively
used in studying quantum complexity, partly because it is a
natural quantum analog of classical decision tree complexity, and
partly because many known quantum algorithms fall into this
framework, including Simon's algorithm \cite{Si97}, Shor's period
finding \cite{Sh97}, Grover's searching algorithm \cite{Gr96} and
many others \cite{BCW98, BDH+01, Da98, DJ92, HNS02}. In the query
model, the input is accessed by querying an oracle, and the goal
is to minimize the number of queries made. We are most interested
in double sided bound-error computation, where the output is
correct with probability at least 2/3 for all inputs. We use
$Q_2(f)$ to denote minimal number of queries for computing $f$
with double sided bound-error. For more details on quantum query
model, we refer to \cite{Am02, BW02} as excellent surveys.

Two main lower bound techniques for $Q_2(f)$ are the polynomial
method \cite {BBC+01} and Ambainis's lower bounds \cite{Am00}, the
latter of which is also called quantum adversary method. Many
lower bounds have recently been achieved by applying the
polynomial method \cite{Aa02, BBC+01, NW99, Ra02, Sh02} and
Ambainis's lower bounds\cite{Aa03, Am00, Am03, CMY03, YZ03}.
Recently, Aaronson even used Ambainis's lower bound technique to
achieve lower bounds for some classical algorithms \cite{Aa03}.
Given the usefulness of the two methods, it is interesting to know
how tight they are. In a recent work \cite{Am03}, Ambainis proved
that polynomial method is not tight, by showing a function with
polynomial degree $M$ and quantum query complexity
$\Omega(M^{1.321...})$. So a natural question is the power of
Ambainis's lower bounds. We show that all known Ambainis's lower
bounds are not tight either, among other results.

There are several known versions of Ambainis's lower bounds, among
which the three Ambainis's theorems are widely used partly because
they have simple forms and are thus easy to use. The first
\emph{Alb}'s two are given in \cite{Am00} as follows.
\begin{Thm} [Ambainis, \cite{Am00}]
Let $f: \B^N \rightarrow \B$ be a function and $X,Y$ be two sets
of inputs \st $f(x) \neq f(y)$ if $x\in X$ and $y\in Y$. Let $R
\subseteq X \times Y$ be a relation \st
\begin{enumerate}
\item $\forall x \in X$, there are at least $m$ different $y \in
Y$ \st $(x,y) \in R$. \vspace{-.3em}

\item $\forall y \in Y$, there are at least $m'$ different $x \in
X$ \st $(x,y) \in R$. \vspace{-.3em}

\item $\forall x \in X$, $\forall i \in [N]$, there are at most
$l$ different $y \in Y$ \st $(x,y) \in R$ and $x_i \neq y_i$.
\vspace{-.3em}

\item $\forall y \in Y$, $\forall i \in [N]$, there are at most
$l'$ different $x \in X$ \st $(x,y) \in R$ and $x_i \neq y_i$.
\vspace{-.3em}
\end{enumerate}
Then $Q_2(f) = \Omega(\sqrt{\frac{mm'}{ll'}})$.
\end{Thm}

\begin{Thm} [Ambainis, \cite{Am00}]
Let $f: I^N\rightarrow \B$ be a Boolean function where $I$ is a
finite set, and $X,Y$ be two sets of inputs \st $f(x) \neq f(y)$
if $x\in X$ and $y\in Y$. Let $R \subseteq X \times Y$ satisfy
\begin{enumerate}
\item $\forall x \in X$, there are at least $m$ different $y \in
Y$ \st $(x,y) \in R$. \vspace{-.3em}

\item $\forall y \in Y$, there are at least $m'$ different $x \in
X$ \st $(x,y) \in R$. \vspace{-.3em}
\end{enumerate}
Denote \[l_{x,i} = |\{y: (x,y)\in R, x_i \neq y_i\}|, \qquad
l_{y,i} = |\{x: (x,y)\in R, x_i \neq y_i\}|\] \[l_{max} =
\max_{x,y,i: (x,y)\in R, i\in [N], x_i \neq y_i} l_{x,i}l_{y,i}.\]
Then $Q_2(f) = \Omega(\sqrt{\frac{mm'}{l_{max}}})$.
\end{Thm}

Obviously, Theorem 2 generalizes Theorem 1. In \cite{Am03},
Ambainis gave another (weighted) way to generalize Theorem 1. We
restate it in a form similar to Theorem 1.

\begin{Def}
Let $f: I^N\rightarrow \B$ be a Boolean function where $I$ is a
finite set. Let $X,Y$ be two sets of inputs \st $f(x) \neq f(y)$
if $x\in X$ and $y\in Y$. Let $R\subseteq X\times Y$ be a
relation. A weight scheme for $X, Y, R$ consists three weight
functions $w(x,y)>0$, $u(x,y,i)>0$ and $v(x,y,i)>0$ satisfying
\begin{equation}
u(x,y,i)v(x,y,i) \geq w^2(x,y) \label{eq:r1}
\end{equation} for
all $(x,y)\in R$ and $i\in [N]$ with $x_i\neq y_i$. We further
denote
\[
w_x = \sum_{y: (x,y)\in R} w(x,y), \qquad w_y = \sum_{x: (x,y)\in
R} w(x,y)
\]
\[
u_{x,i} = \sum_{y:(x,y)\in R, x_i \neq y_i} u(x,y,i),\quad v_{y,i}
= \sum_{x:(x,y)\in R, x_i \neq y_i} v(x,y,i).
\]
\end{Def}

\begin{Thm} [Ambainis, \cite{Am03}]
Let $f: I^N\rightarrow \B$ where $I$ is a finite set, and
$X\subseteq f^{-1}(0)$, $Y\subseteq f^{-1}(1)$ and $R\subseteq
X\times Y$. Let $w,u,v$ be a weight scheme for $X, Y, R$. Then
\[Q_2(f) =\Omega(\sqrt{\min_{x\in X,i\in [N]}\frac{w_x}{u_{x,i}}
\cdot \min_{y\in Y,j\in [N]}\frac{w_y}{v_{y,j}}})\]
\end{Thm}

\vspace{1em} Let us denote by $Alb_1(f)$, $Alb_2(f)$ and
$Alb_3(f)$ the best lower bound for function $f$ achieved by
Theorem 1, 2 and 3, respectively\footnote{To make the later
results more precise, we actually use $Alb_i(f)$ to denote the
value inside the $\Omega()$ notation. For example, $Alb_1(f) =
\max_{(X,Y,R)}\sqrt{\frac{mm'}{ll'}}$.}. Note that in the four
$Alb$'s, there are many parameters $(X, Y, R, u, v, w)$ to be set.
By setting these parameters in an appropriate way, one can get
lower bounds of quantum query complexity for many problems. In
particular, we consider the following three graph
properties\footnote{In this paper, all the graph property problems
are given by adjacency matrix input.}.
\begin{enumerate}
\item \textsc{Bipartiteness}: \emph{Given an undirected graph,
decide whether it is a bipartite graph.}

\item \textsc{Graph Matching}: \emph{Given an undirected graph,
decide whether it has a perfect matching.}

\item \textsc{Bipartite Matching}: \emph{Given an undirected
bipartite graph, decide whether it has a perfect matching.}
\end{enumerate}
We show by using $Alb_2$ that all these three graph properties
have a $\Omega(n^{1.5})$ lower bound, where $n$ is the number of
vertices. For \textsc{Bipartiteness}, this improves the previous
result of $\Omega(n)$ lower bound by Laplante and Magniez
\cite{LM03} and Durr \etal \cite{DHH+03}.

Since $Alb_2$ and $Alb_3$ generalizes $Alb_1$ in different ways,
it is interesting to compare them and see which is more powerful.
It turns out that $Alb_2(f) \leq Alb_3(f)$.

However, even $Alb_3$ has a limitation: we show that $Alb_3(f)
\leq \sqrt{N\cdot \min\{C_0(f), C_1(f)\}}$, where $C_0(f)$ and
$C_1(f)$ are the 0-and 1-certificate complexity, respectively.
This has two immediate consequences. First, it gives a negative
answer to the open problem whether $Alb_2$ or $Alb_3$ is tight,
because for \textsc{Element Distinctness}, we know that $Q_2(f) =
\Theta(N^{2/3})$ while on the other hand we have $\sqrt{N\cdot
\min\{C_0(f), C_1(f)\}} = \sqrt{2N}$.

Second, for some problems whose precise quantum query complexities
are still unknown, our theorem implies that the best known lower
bound cannot be further improved by using Ambainis's lower bound
techniques, no matter how we choose the parameters in the $Alb$
theorems. For example \textsc{Triangle}/$k$-\textsc{Clique} ($k$
is constant) are the problems to decide whether an $n$-node graph
contains a triangle/$k$-node clique. It is easy to get a
$\Omega(n)$ lower bound for both of them, and by our theorem, this
is the best possible by using Ambainis's lower bound techniques.
Also the $\Omega(n^{1.5})$ lower bound for \textsc{Bipartiteness},
\textsc{Bipartite Matching} and \textsc{Graph Matching} cannot be
further improved by $Alb$'s because the $C_1(f) = O(n)$ for all of
them.

Further, if $f$ is a total function, then the above upper bound
for $Alb$'s can be further tightened in two ways. The first one is
that $Alb_3(f) \leq \sqrt{N \cdot CI(f)}$, where $CI(f)$ is the
the size of the largest intersection of a 0-certificate set and a
1-certificate set, so $CI(f) \leq C_-(f)$. The second approach
leads to another result $Alb_3(f) \leq \sqrt{C_0(f)C_1(f)}$. Both
the results imply that for \textsc{And-Or Tree}, a problem whose
quantum query complexity is still open \cite{Am03}, the current
best $\Omega(\sqrt{N})$ lower bound cannot be further improved by
using Ambainis's lower bounds.

It is also natural to consider combining the different approaches
that $Alb_2$ and $Alb_3$ use to generalize $Alb_1$, and get a
further general one. Based on this idea, we give a new and more
generalized lower bound theorem, which we call $Alb_4$. Compared
with $Alb_3$, this may be easier to use.

\vspace{1em} \noindent \textbf{Related work}

Recently, Szegedy independently shows that $Alb_3(f) \leq
\sqrt{N\cdot C_-(f)}$ in general and $Alb_3(f) \leq
\sqrt{C_0(f)C_1(f)}$ in a different way \cite{Sz04}. He also shows
in \cite{Sz04} that $Alb_3$ by Ambainis \cite{Am03}, $Alb_4$ in
the present paper \cite{Zh03}, and another quantum adversary
method proposed in \cite{BSS03} are equivalent.

The theorem $Alb_3(f) \leq \sqrt{N\cdot C_-(f)}$ is also obtained
by Laplante and Magniez by using Kolmogorov complexity
\cite{LM03}.

\section{Old Ambainis's lower bounds}
In this section we first use $Alb_2$ to derive $\Omega(n^{1.5})$
lower bounds for \textsc{Bipartiteness}, \textsc{Bipartite
Matching} and \textsc{Graph Matching}, then show that $Alb_3$ has
actually at least the same power as $Alb_2$.

\begin{Thm}
All the three graph properties \textsc{Bipartiteness},
\textsc{Bipartite Matching} and \textsc{Graph Matching} have
$Q_2(f) = \Omega(n^{1.5})$.
\end{Thm}
\begin{proof}
1. \textsc{Bipartiteness}. The proof is very similar to the one
for proving $\Omega(n^{1.5})$ lower bound of \textsc{Graph
Connectivity} by Durr \etal \cite{CMY03}. Without loss of
generality, we assume $n$ is even, because otherwise we can use
the following argument on arbitrary $n-1$ (out of total $n$) nodes
and leave the $n^{th}$ node isolated. Let

$X = \{G: G$ is composed of a single n-length cycle\},

$Y = \{G: G$ is composed of two cycles each with length being an
odd number between $n/3$ and $2n/3$\}, and

$R = \{(G,G')\in X\times Y: \exists$ four nodes $v_1, v_2, v_3,
v_4$ \st the only difference between graphs $G$ and $G'$ is that
$(v_1, v_2), (v_3, v_4)$ are edges in $G$ but not in $G'$ and
$(v_1, v_3), (v_2, v_4)$ are edges in $G'$ but not in $G$\}.

Note that a graph is bipartite if and only if it contains no cycle
with odd length. Therefore, any graph in $X$ is a bipartite graph
because $n$ is even, and any graph in $Y$ is not bipartite graph
because it contains two odd-length cycles. Then all the remaining
analysis is the same as calculation in the proof for \textsc{Graph
Connectivity} (undirect graph and matrix input) in \cite{CMY03},
and finally $Alb_2(\textsc{Bipartiteness}) = \Omega(n^{1.5})$.

2. \textsc{Bipartite Matching}. Let $X$ be the set of the
bipartite graphs like Figure 1(a) where $\tau$ and $\sigma$ are
two permutations of $\{1, ..., n\}$, and $\frac{n}{3} \leq k \leq
\frac{2n}{3}$. Let $Y$ be the set of the bipartite graphs like
Figure 1(b), where $\tau'$ and $\sigma'$ are two permutations of
$\{1, ..., n\}$, and also $\frac{n}{3} \leq k' \leq \frac{2n}{3}$.
It is easy to see that all graphs in $X$ have no matching, while
all graphs in $Y$ have one.
\begin{figure}[h]
\begin{center}
\epsfig{file=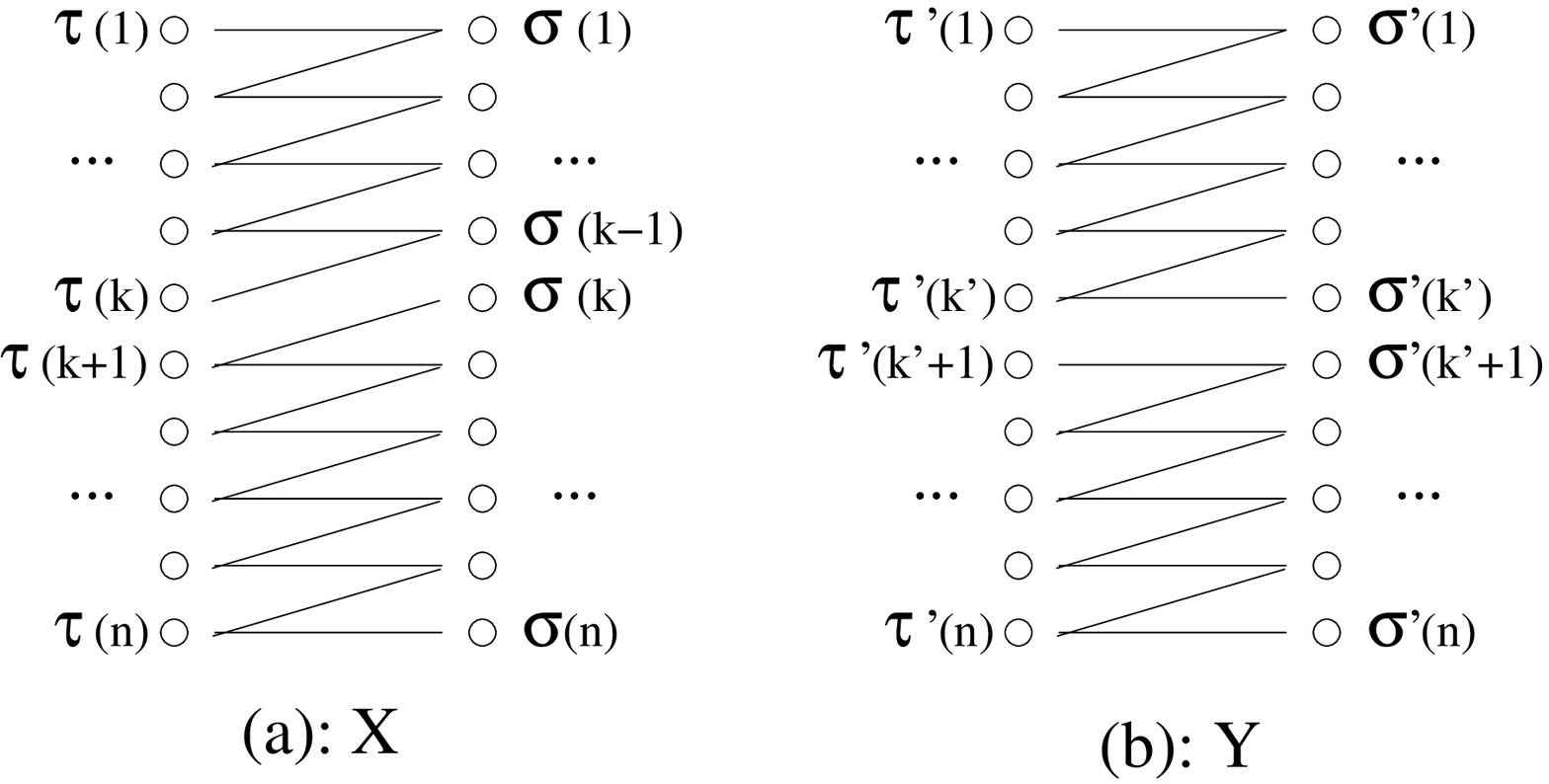, width=9.8cm} \vspace{-1em} \caption{$X$
and $Y$}
\end{center}
\end{figure}

Let $R$ be the set of all pairs of $(x,y)\in X\times Y$ as in
Figure 2, where graph $y$ is obtained from $x$ by choosing two
horizonal edges $(\tau(i), \sigma(i))$, $(\tau(j), \sigma(j))$,
removing them, and adding two edges $(\tau(i), \sigma(j))$,
$(\tau(j), \sigma(i))$.
\begin{figure}[h]
\begin{center}
\epsfig{file=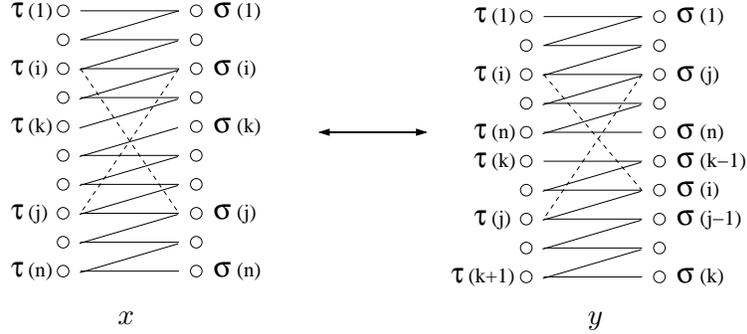, width=9.8cm}\\
\hspace{-1em}$x$ \hspace{15em} $y$ \vspace{-0.5em}
\caption{$R\subseteq X\times Y$}
\end{center}
\end{figure}

Now it is not hard to calculate the $m, m', l_{max}$ in $Alb_2$.
For example, to get $m$ we study $x$ in two cases. When
$\frac{n}{3} \leq k \leq \frac{n}{2}$, any edge $(\tau(i),
\sigma(i))$ where $i\in [k-n/3, k]$ has at least $n/6$ choices for
edge $(\tau(j), \sigma(j))$ because the only requirement for
choosing is that $k'\in [n/3, 2n/3]$ and $k' = j-i$. The case when
$\frac{n}{2} \leq k \leq \frac{2n}{3}$ can be handled
symmetrically. Thus $m = \Theta(n^2)$. Same argument yields
$m'=\Theta(n^2)$. Finally, for $l_{max}$, we note that if the edge
$e = (\tau(i), \sigma(i))$ for some $i$, then $l_{x,e} = O(n)$ and
$l_{y,e} = 1$; if the edge $e = (\tau(i), \sigma(j))$ for some
$i,j$, then $l_{x,e} = 1$ and $l_{y,e} = O(n)$. For all other
edges $e$, $l_{x,e} = l_{y,e} = 0$. Putting all together, we have
$l_{max} = O(n)$. Thus by Theorem 2, we know that
$Alb_2(\textsc{Bipartite Matching}) = \Omega(n^{1.5})$.

3. \textsc{Graph Matching}. This can be easily shown either by
using the same $(X, Y, R)$ as the proof for
\textsc{Bipartiteness}, because a cycle with odd length has no
matching, or by noting that \textsc{Bipartite Matching} is a
special case of \textsc{Graph Matching}.
\end{proof}

It is interesting to note that we can also prove the above theorem
by $Alb_3$. For example, for \textsc{Bipartite Matching}, We
choose $X, Y, R$ in the same way, and let $w(x,y) = 1$ for all
$(x,y)\in R$. Let $u(x,y,e) = 1/\sqrt{n}$ if $e$ is a horizonal
edge $(\tau(i), \sigma(i))$ in $x$, and $u(x,y,e) = \sqrt{n}$ if
$e = (\tau(i), \sigma(j))$ or $e = (\tau(j), \sigma(i))$ in $x$.
Thus $u_{x,e} = \Theta(\sqrt{n})$ for all edge $e$, it is the same
for $v_{y,e}$, thus $w_x/u_{x,e} = \Theta(n^{1.5}), w_y/v_{y,e} =
\Theta(n^{1.5})$, and $Q_2(f) = \Omega(n^{1.5})$ by $Alb_3$.

This coincidence is not accident. Actually it turns out that we
can always show a lower bound by $Alb_3$ provided that it can be
shown by $Alb_2$.

\begin{Thm}
$Alb_2(f) \leq Alb_3(f)$.
\end{Thm}
\begin{proof}
For any $X, Y, R$ in Theorem 2, we set the weight functions in
Theorem 3 as follows. Let $w(x,y) = 1$, $u(x,y,i) =
\sqrt{l_{max}}/l_{x,i}$ and $v(x,y,i) = \sqrt{l_{max}}/l_{y,i}$.
It's easy to check that \[u(x,y,i)v(x,y,i) =
\frac{l_{max}}{l_{x,i} l_{y,i}} \geq 1 = w(x,y)\] Now that
$u(x,y,i)$ is independent on $y$, so we have $u_{x,i} =
l_{x,i}u(x,y,i) = \sqrt{l_{max}}$. Symmetrically, it follows that
$v_{y,i} = \sqrt{l_{max}}$. Thus, by denoting $m_x=|\{y: (x,y)\in
R\}|$ and $m_y=|\{x: (x,y)\in R\}|$, we have
\[\min_{x,i} \frac{w_x}{u_{x,i}} \min_{y,i} \frac{w_y}{v_{y,i}} =
\min_{x,i} \frac{m_x}{\sqrt{l_{max}}} \min_{y,i}
\frac{m_y}{\sqrt{l_{max}}} = \frac{m}{\sqrt{l_{max}}}
\frac{m'}{\sqrt{l_{max}}} = \frac{mm'}{l_{max}}\]which means that
for any $X,Y,R$ in Theorem 2, the lower bound result can be also
achieved by Theorem 3.
\end{proof}

\section{Limitations of Ambainis's lower bounds}
In this section, we show some bounds for the $Alb$'s in terms of
certificate complexity. We consider Boolean functions.
\begin{Def}
For an $N$-ary Boolean function $f: I^N \rightarrow \B$ and an
input $x\in I^N$, a certificate set $CS_x$ of $f$ on $x$ is a set
of indices such that $f(x) = f(y)$ whenever $y_i = x_i$ for all
$i\in CS_x$. The certificate complexity $C(f,x)$ of $f$ on $x$ is
the size of a smallest certificate set of $f$ on $x$. The
$b$-certificate complexity of $f$ is $C_b(f) = \max_{x:
f(x)=b}C(f,x)$. The certificate complexity of $f$ is $C(f) =
\max\{C_0(f), C_1(f)\}$. We further denote $C_-(f) = \min\{C_0(f),
C_1(f)\}$.
\end{Def}

\subsection{A general limitation for Ambainis's lower bounds}
In this subsection, we give an upper bound for $Alb_4(f)$, which
implies a limitation of all the three known Ambainis's lower bound
techniques.

\begin{Thm}
$Alb_3(f) \leq \sqrt{N \cdot C_-(f)}$, for any $N$-ary Boolean
function $f$.
\end{Thm}
\begin{proof} Actually we prove a stronger result: for any $(X, Y,
R, u, v, w)$ as in Theorem 3,
\[\min_{(x, y)\in R, i\in [N]}\frac{w_xw_y}{u_{x,i}v_{y,i}} \leq NC_-(f).\]
With out loss of generality, we assume that $C_-(f) = C_0(f)$, and
$X\subseteq f^{-1}(0)$ and $Y\subseteq f^{-1}(1)$. We can actually
further assume that $R = X \times Y$, because otherwise we just
let $R' = X \times Y$, and set new weight functions as follows.
\[u'(x,y,i) =
\begin{cases}
u(x,y,i) & (x,y) \in R \\
0 & otherwise \end{cases}, \qquad v'(x,y,i) =
\begin{cases}
v(x,y,i) & (x,y) \in R \\
0 & otherwise \end{cases},\]
\[ w'(x,y) =
\begin{cases}
w(x,y) & (x,y) \in R \\
0 & otherwise \end{cases}.\] Then it is easy to see that it
satisfies \eqref{eq:r1} so it is also a weight scheme. And for
these new weight functions, we have $u'_{x,i} = \sum_{y: (x,y)\in
R', x_i \neq y_i} u'(x,y,i) = \sum_{y: (x,y)\in R, x_i \neq y_i}
u(x,y,i) = u_{x,i}$ and similarly $v'_{y,i} = v_{y,i}$ and $w'_x =
w_x, w'_y = w_y$.\footnote{Note that the function values of $u',
v', w'$ are zero when $(x,y)\neq R$, which does not conform to the
definition of weight scheme. But actually Theorem 3 also holds for
$u\geq 0, v\geq 0, w\geq 0$ as long as $u_{x,i}, v_{y,i}, w_x,
w_y$ are all strictly positive for any $x,y,i$. This can be seen
from the proof of $Alb_4$ in Section 4.} It follows that
$\frac{w_xw_y}{u_{x,i}v_{y,i}} =
\frac{w'_xw'_y}{u'_{x,i}v'_{y,i}}$, thus we can use $(X', Y', R',
u', v', w')$ to derive the same lower bound as we use $(X, Y, R,
u, v, w)$.

So now we suppose $R = X \times Y$ and We prove that $\exists x\in
X, y\in Y, i\in [N]$, \st
\[
w_x w_y \leq N \cdot C_0(f)\ u_{x,i} v_{y,i},
\]
Suppose the claim is not true. Then for all $x\in X, y\in Y, i\in
[N]$, we have
\begin{equation}
w_x w_y > N \cdot C_0(f)\ u_{x,i} v_{y,i}. \label{eq:r2}
\end{equation}
We first fix $i$ for the moment. And for each $x\in X$, we fix a
smallest certificate set $CS_x$ of $f$ on $x$. Clearly $|CS_x|
\leq C_0(f)$. We sum \eqref{eq:r2} over $\{x\in X: i \in CS_x\}$
and $\{y \in Y\}$. Then we get
\begin{equation}
\sum_{x\in X:\ i\in CS_x} w_x \sum_{y\in Y} w_y > N \cdot C_0(f)
\sum_{x\in X:\ i\in CS_x} u_{x,i} \sum_{y\in Y} v_{y,i}.
\label{eq:r3}
\end{equation}
Note that $\sum_{y\in Y} w_y = \sum_{x\in X, y\in Y} w(x,y) =
\sum_{x\in X} w_x$, and that $\sum_{y\in Y} v_{y,i} = \sum_{x\in
X, y\in Y: x_i \neq y_i} v(x,y,i) = \sum_{x\in X} v_{x,i}$ where
$v_{x,i} = \sum_{y\in Y: x_i \neq y_i} v(x,y,i)$. Inequality
\eqref{eq:r3} turns to
\[
\begin{array}{rcl}
\sum_{x\in X:\ i\in CS_x} w_x \sum_{x\in X} w_x & > & N \cdot
C_0(f)
\sum_{x\in X:\ i\in CS_x} u_{x,i} \sum_{x\in X} v_{x,i} \\
& \geq & N \cdot C_0(f) \sum_{x\in X:\ i\in CS_x} u_{x,i}
\sum_{x\in X:\
i\in CS_x} v_{x,i} \\
& \geq & N \cdot C_0(f) (\sum_{x\in X:\ i\in CS_x} \sqrt{u_{x,i}
v_{x,i}})^2
\end{array}
\]
due to Cauchy-Schwartz Inequality. We further note that
\[
\begin{array}{rcl}
u_{x,i}v_{x,i} & = & \sum_{y\in Y: x_i \neq y_i} u(x,y,i)
\sum_{y\in Y: x_i \neq y_i} v(x,y,i) \\
& \geq & (\sum_{y\in Y: x_i \neq y_i}
\sqrt{u(x,y,i)v(x,y,i)})^2 \\
& \geq & (\sum_{y\in Y: x_i \neq y_i} w(x,y))^2 \\
& = & (w_{x,i})^2
\end{array}
\]
where we define $w_{x,i} = \sum_{y\in Y: x_i \neq y_i} w(x,y)$.
Thus
\begin{equation}
\sum_{x\in X:\ i\in CS_x} w_x \sum_{x\in X} w_x > N \cdot C_0(f)
(\sum_{x\in X:\ i\in CS_x} w_{x,i})^2 \label {eq:r4}
\end{equation}

Now we sum \eqref{eq:r4} over $i=1,...,N$, and note that
\[
\sum_i \sum_{x\in X:\ i\in CS_x} w_x = \sum_{x\in X} \sum_{i: i\in
CS_x} w_x \leq C_0(f) \sum_{x\in X} w_x
\]
because $|CS_x| \leq C_0(f)$ for each $x$. We have
\[
(\sum_{x\in X} w_x)^2  >  N \sum_{i=1}^N (\sum_{x\in X:\ i\in
CS_x} w_{x,i})^2
\]
By the arithmetic-square average inequality $N(a_1^2 + ... +
a_N^2) \geq (a_1 + ... + a_N)^2$, we have
\[
(\sum_{x\in X} w_x)^2  >  (\sum_{x\in X, i\in [N]:\ i\in CS_x}
w_{x,i})^2
 = (\sum_{x\in X, i\in [N], y\in Y:\ i\in CS_x, x_i \neq y_i}
 w(x,y))^2\] \[= (\sum_{x\in X, y\in Y} \sum_{\ i\in [N]:\ i\in CS_x, x_i \neq y_i}
 w(x,y))^2
\]
But by the definition of certificate, we know that for any $x$ and
$y$ there is at least one index $i \in CS_x$ \st $x_i \neq y_i$.
Therefore, we derive an inequality
\[(\sum_{x\in X} w_x)^2
> (\sum_{x\in X, y\in Y} w(x, y))^2 = (\sum_{x\in X} w_x)^2\] which is a contradiction, as desired.
\end{proof}

We add some comments about this upper bound of $Alb_3$. First,
this bound looks weak at first glance because the $\sqrt{N}$
factor seems too large. But in fact it is necessary at least for
partial functions. Consider the problem of \textsc{Invert A
Permutation}\cite{Am00}\footnote{The original problem is not a
Boolean function, but we can define a Boolean-valued version of
it. Instead of finding the position $i$ with $x_i=1$, we are to
decide whether $i$ is odd or even. The original proof of the
$\Omega(\sqrt{N})$ lower bound still holds. }, where $C_0(f) =
C_1(f) = 1$ but even the $Alb_2(f) = \Omega(\sqrt{N})$.

Second, the quantum query complexity of \textsc{Element
Distinctness} is known to be $\Theta(N^{2/3})$. The lower bound
part is obtained by Shi \cite{Sh02} (for large range) and Ambainis
\cite{Am03c} (for small range); the upper bound part is obtained
by Ambainis \cite{Am03b}. Observe that $C_1(f) = 2$ thus $\sqrt{N
C_1(f)} = \Theta(N)$, we derive the following interesting
corollary from the above theorem.
\begin{Cor}
$Alb_3$ is not tight.
\end{Cor}

\vspace{1em} We make some remarks on the quantity $\sqrt{N\cdot
C_-(f)}$ to end this subsection. A function $f$ is symmetric if
$f(x_1...x_N) = f(x_{\sigma(1)}...x_{\sigma(n)})$ for any input
$x$ and any permutation $\sigma$ on $[N]$. In \cite{BBC+01}, Beal
\etal prove that $Q_2(f) = \Theta(\sqrt{N(N-\Gamma(f))})$ by using
Paturi's result $\widetilde{deg}(f) =
\Theta(\sqrt{N(N-\Gamma(f))})$ \cite{Pa92}, where $\Gamma(f) =
\min\{|2k-n+1|: f_k \neq k_{k+1}, 0\leq k \leq n-1\}$. It is not
hard to show that $\Gamma(f) = N - \Theta(C_-(f))$ for symmetric
function $f$. Thus we know that both $\widetilde{deg}(f)$ and
$Q_2(f)$ can also be described in terms of certificate complexity
as $\Theta(\sqrt{N\cdot C_-(f)})$.

\subsection{Two better upper bounds for total functions}
It turns out that if the function is total, then the upper bound
can be further tightened. We introduce a new measure which
basically characterizes the size of intersection of a 0- and
1-certificate sets.
\begin{Def}
For any function $f$, if there is a certificate set assignment
$CS: \B^N \rightarrow 2^{[N]}$ such that for any inputs $x,y$ with
$f(x) \neq f(y)$, $|CS_x \cap CS_y| \leq k$, then $k$ is called a
candidate certificate intersection complexity of $f$. The minimal
candidate certificate intersection complexity of $f$ is called the
certificate intersection complexity of $f$, denoted by $CI(f)$. In
other words, $CI(f) = \min_{CS}\max_{x,y: f(x)\neq f(y)} |CS_x\cap
CS_y|$.
\end{Def}
Now the improved theorem is as follows. Note that by the above
definition we know $CI(f) \leq C_-(f)$, thus the the following
theorem really improves Theorem 7 for total functions.

\begin{Thm}
$Alb_3(f) \leq \sqrt{N \cdot CI(f)}$, for any $N$-ary total
Boolean function $f$.
\end{Thm}
\begin{proof}
Again, we prove a stronger result that for any $(X, Y, R, u, v,
w)$ in Theorem 3, \[\min_{(x,y)\in R, i\in [N]}
\frac{w_xw_y}{u_{x,i}v_{y,i}}\leq N\cdot CI(f).\] Similar to the
proof for Theorem 6, we assume with out loss of generality that $R
= X \times Y$ and for all $x\in X, y\in Y$, we have
\begin{equation} w_x w_y > N\cdot CI(f)\ u_{x,i} v_{y,i}.
\label{eq:r5}
\end{equation}
and we shall show a contradiction. Now first fix $i$ and sum
\eqref{eq:r5} over $\{x\in X: i\in CS_x\}$ and $\{y\in Y: i\in
CS_y\}$. We get
\[
\begin{array}{rcl}
\sum_{x\in X,y\in Y:\ i\in CS_x\cap CS_y} w_x w_y & > & N\cdot
CI(f)\ \sum_{x\in X:\ i\in
CS_x} u_{x,i} \sum_{y\in Y:\ i\in CS_y} v_{y,i} \\
& = & N\cdot CI(f)\ \sum_{x\in X,y\in Y:\ i\in CS_x, x_i \neq y_i}
u(x,y,i)\vspace {.2em} \\
\vspace {.5em} & & \hspace{4.5em} \cdot \sum_{x\in X,y\in Y:\ i\in CS_y, x_i \neq y_i} v(x,y,i) \\
& \geq & N\cdot CI(f)\ \sum_{x\in X,y\in Y:\ i\in CS_x\cap CS_y,
x_i \neq
y_i} u(x,y,i) \vspace{.2em} \\
\vspace{.5em} & & \hspace{4.5em} \cdot \sum_{x\in X,y\in Y:\ i\in CS_x\cap CS_y, x_i \neq y_i} v(x,y,i) \\
& \geq & N\cdot CI(f)\ (\sum_{x\in X,y\in Y:\ i\in CS_x\cap CS_y,
x_i \neq
y_i} \sqrt{u(x,y,i) v(x,y,i)})^2 \\
& \geq & N\cdot CI(f)\ (\sum_{x\in X,y\in Y:\ i\in CS_x\cap CS_y,
x_i \neq y_i} w(x,y))^2
\end{array}
\]
Now sum over $i=1, ..., N$, we get
\[
\begin{array}{rcl}
\sum_{x\in X,y\in Y,i\in [N]:\ i\in CS_x\cap CS_y} w_x w_y & > &
N\cdot CI(f)\ \sum_{i=1}^N (\sum_{x\in X,y\in Y:\ i\in CS_x\cap
CS_y, x_i \neq y_i}
w(x,y))^2 \\
& \geq & CI(f)\  (\sum_{x\in X,y\in Y,i\in [N]:\ i\in CS_x\cap
CS_y, x_i \neq y_i} w(x,y))^2
\end{array}
\]
Note that for total function $f$, if $f(x) \neq f(y)$, there is at
least one position $i\in CS_x \cap CS_y$ \st $x_i \neq y_i$. Thus
\[
\sum_{x\in X,y\in Y,i\in [N]:\ i\in CS_x\cap CS_y, x_i \neq y_i}
w(x,y) \geq \sum_{x\in X, y\in Y} w(x,y)
\]
On the other hand, by the definition of $CI(f)$, we have
\[
\sum_{x\in X,y\in Y,i\in [N]:\ i\in CS_x\cap CS_y} w_x w_y \leq
CI(f) \sum_{x\in X,y\in Y} w_x w_y =  CI(f) (\sum_{x\in X,y\in
Y}w(x,y))^2
\]
Therefore we get a contradiction \[CI(f) (\sum_{x\in X,y\in
Y}w(x,y))^2 > CI(f) (\sum_{x\in X,y\in Y}w(x,y))^2,\] as desired.
\end{proof}

In \cite{Am03}, Ambainis proposed the open problem \textsc{And-Or
Tree}. In the problem, there is a complete binary tree with height
$2n$. Any node in odd levels is labelled with AND and any node in
even levels is labelled with OR. The $N=4^n$ leaves are the input
variables, and the value of the function is the value that we get
at the root, with value of each internal node calculated from the
values of its two children in the common AND/OR interpretation.
The best quantum lower bound is $\Omega(\sqrt{N})$ and best
quantum upper bound is no more than the best classical
(randomized) one $O(N^{0.753...}) = O((\frac{1+\sqrt{33}}{4})^n)$.
Note that $C_-(\textsc{And-Or Tree}) = 2^n = \sqrt{N}$ and thus
$\sqrt{NC_-(f)} = N^{3/4}$. Using the above theorem, we know that
we cannot improve best known lower bounds of \textsc{And-Or Tree}
by $Alb$'s.

\begin{Cor}
$Alb_4(\textsc{And-Or Tree}) \leq \sqrt{N}$.
\end{Cor}
\vspace{-0.5em}\begin{proof} It is sufficient to prove that there
is a certificate assignment $CS$ \st $|CS_x\cap CS_y|=1$ for any
$f(x)\neq f(y)$. In fact, by a simple induction, we can prove that
the standard certificate assignment satisfies this property. The
base case is trivial. For the induction step, we note that for an
AND connection of two subtrees, the 0-certificate set of the new
larger tree can be chosen as any one of the two 0-certificate sets
of the two subtrees, and the 1-certificate set of the new larger
tree can be chosen as the union of the two 1-certificate sets of
the two subtrees. As a result, the intersection of the two new
certificate sets is not enlarged. The OR connection of two
subtrees is analyzed in the same way. Thus the intersection of the
final 0- and 1-certificate sets is of size 1.
\end{proof}

We can also tighten the $\sqrt{N\cdot C_-(f)}$ upper bound in
another way and get the following result, which also implies
Corollary 9. 
\begin{Thm}
$Alb_3(f) \leq \sqrt{C_0(f)C_1(f)}$, for any total Boolean
function $f$.
\end{Thm}
\begin{proof}
For any $(X, Y, R, u, v, w)$ in Theorem 3, we assume without loss
of generality that $X\subseteq f^{-1}(0), Y\subseteq f^{-1}(1)$
and $R = X\times Y$. We are to prove $\exists x, y, i, j$ \st
$w_xw_y\leq C_0(f)C_1(f)u_{x,i}v_{y,j}$. Suppose this is not true,
\ie for all $x\in X, y\in Y, i,j\in [N]$, $w_xw_y >
C_0(f)C_1(f)u_{x,i}v_{y,j}$. First fix $x,y$ and sum over $i\in
CS_x$ and $j\in CS_y$. Since $|CS_x|\leq C_0(f), |CS_y|\leq
C_1(f)$, we have
\[
w_xw_y > \sum_{i\in CS_x} u_{x,i} \sum_{j\in CS_y} v_{y,j}\\
\]
Now we sum over $x\in X$ and $y\in Y$,
\[
\begin{array}{rcl}
\sum_{x\in X}w_x\sum_{y\in Y}w_y & > & \sum_{x\in X, i\in CS_x} u_{x,i} \sum_{y\in Y, j\in CS_y} v_{y,j}\\
& = & \sum_{x\in X,y\in Y,i\in [N]: x_i \neq y_i, i \in
CS_x}u(x,y,i) \sum_{x\in X,y\in Y,j\in [N]: x_j \neq y_j, j \in
CS_y}v(x,y,j) \\
\end{array}
\]
Since $f$ is total, there is at least one $i_0\in CS_x\cap CS_y$
\st $x_{i_0}\neq y_{i_0}$. Thus
\[
\begin{array}{rcl}
\sum_{x\in X}w_x\sum_{y\in Y}w_y & > & \sum_{x\in X,y\in Y}
u(x,y,i_0) \sum_{x\in X,y\in Y}v(x,y,i_0) \\
 & \geq & \sum_{x\in X,y\in Y} u(x,y,i_0) v(x,y,i_0) \\
 & \geq & (\sum_{x\in X,y\in Y} \sqrt{u(x,y,i_0) v(x,y,i_0) })^2
 \\
 & \geq & (\sum_{x\in X,y\in Y} w(x,y))^2 \\
 & = & \sum_{x\in X}w_x\sum_{y\in Y}w_y
\end{array}
\]
which is a contradiction.
\end{proof}

Finally, we remark that these two improved upper bounds of
$Alb_3(f)$ are not always tight. In \cite{YZ03}, Yao and Zhang
prove that two graph properties \textsc{Scorpion} and
\textsc{Sink} both have $Q_2(f) = \Theta(\sqrt{n})$. But both
$\sqrt{C_0(f)C_1(f)}$ and $\sqrt{N\cdot CI(f)}$ are $\Theta(n)$.

\section{A further generalized Ambainis's lower bound}
While $Alb_2$ and $Alb_3$ use different ideas to generalize
$Alb_1$, it is natural to combine both and get a further
generalization. The following theorem is a result in this
direction. This theorem to Theorem 3 is as Theorem 2 to Theorem 1.
The proof is similar to the ones in \cite{Am00, Am03}, with inner
products substituted for density operators to make it look
easier\footnote{This idea was also used in some other papers such
as \cite{HNS02}.}.

\begin{Thm} Let $f: I^N\rightarrow \B$ where $I$ is a finite set, and
$X,Y$ be two sets of inputs \st $f(x) \neq f(y)$ if $x\in X$ and
$y\in Y$. Let $R \subseteq X \times Y$. Let $w,u,v$ be a weight
scheme for $X, Y, R$. Then
\[Q_2(f) =\Omega(\sqrt{\min_{(x,y)\in R,i\in [N], x_i\neq y_i}
\frac{w_xw_y}{u_{x,i}v_{y,i}}})\]
\end{Thm}
\begin{proof} The query computation is a sequence of operations
$U_0 \rightarrow O_x \rightarrow U_1 \rightarrow ... \rightarrow
U_T$ on some fixed initial state, say $\ket{0}$. Note that here
$T$ is the number of queries. Denote $\ket{\psi_x^k} =
U_{k-1}O_x...U_1O_xU_0 \ket{0}$. Note that $\ket{\psi_x^1} =
\ket{0}$ for all input $x$. Because the computation is correct
with high probability ($1-\epsilon$), for any $(x,y)\in R$, the
two final states have to have some distance to let the measurement
distinguish them. In other words, we can assume that
$|\qip{\psi_x^T}{\psi_y^T}| \leq c$ for some constant $c<1$. Now
suppose that
\[\ket{\psi_x^{k-1}} = \sum_{i,a,z}
\alpha_{i,a,z} \ket{i,a,z},\ \ \ket{\psi_y^{k-1}} = \sum_{i,a,z}
\beta_{i,a,z} \ket{i,a,z}\] where $i$ is for the index address,
$a$ is for the answer, and $z$ is the workspace. Then the oracle
works as follows.
\[O_x\ket{\psi_x^{k-1}} = \sum_{i,a,z} \alpha_{i,a,z} \ket{i,a\oplus x_i,z}
= \sum_{i,a,z} \alpha_{i,a\oplus x_i,z} \ket{i,a,z}\]
\[O_y\ket{\psi_y^{k-1}} = \sum_{i,a,z} \beta_{i,a,z} \ket{i,a\oplus y_i,z}
= \sum_{i,a,z} \beta_{i,a\oplus y_i,z} \ket{i,a,z}\] So we have
\[
\begin{array}{rcl}
\qip{\psi_x^k}{\psi_y^k} & = & \sum_{i,a,z} \alpha_{i,a\oplus
x_i,z}^* \beta_{i,a\oplus y_i,z} \\
& = & \sum_{i,a,z: x_i=y_i} \alpha_{i,a\oplus x_i,z}^*
\beta_{i,a\oplus y_i,z} + \sum_{i,a,z: x_i \neq y_i}
\alpha_{i,a\oplus x_i,z}^* \beta_{i,a\oplus y_i,z} \\
& = & \qip{\psi_x^{k-1}}{\psi_y^{k-1}} + \sum_{i,a,z: x_i \neq
y_i} \alpha_{i,a\oplus x_i,z}^* \beta_{i,a\oplus y_i,z} -
\sum_{i,a,z: x_i \neq y_i} \alpha_{i,a,z}^* \beta_{i,a,z}
\end{array}
\]
Thus
\[
\begin{array}{rcl}
1-c & = & 1-|\qip{\psi_x^T}{\psi_y^T}| =
\sum_{k=1}^T(|\qip{\psi_x^{k-1}}{\psi_y^{k-1}}|-|\qip{\psi_x^k}{\psi_y^k}|)
\\
& \leq & \sum_{k=1}^T |\qip{\psi_x^{k-1}}{\psi_y^{k-1}} -
\qip{\psi_x^k}{\psi_y^k}| \\
& = & \sum_{k=1}^T |\sum_{i,a,z: x_i \neq y_i} (\alpha_{i,a\oplus
x_i,z}^* \beta_{i,a\oplus y_i,z} -
\alpha_{i,a,z}^* \beta_{i,a,z}) | \\
& \leq & \sum_{k=1}^T \sum_{i,a,z: x_i \neq y_i}
(|\alpha_{i,a\oplus x_i,z}| |\beta_{i,a\oplus y_i,z}| +
|\alpha_{i,a,z}| |\beta_{i,a,z}|)
\end{array}
\]
Summing up the inequalities for all $(x,y)\in R$, with weight
$w(x,y)$ multiplied, yields
\[
\begin{array}{rl}
& (1-c) \sum_{(x,y)\in R}w(x,y) \\
\leq & \sum_{k=1}^T \sum_{(x,y)\in R} \sum_{i,a,z: x_i \neq y_i}
w(x,y) (|\alpha_{i,a\oplus x_i,z}| |\beta_{i,a\oplus y_i,z}| + |\alpha_{i,a,z}| |\beta_{i,a,z}|) \\
\leq & \sum_{k=1}^T \sum_{(x,y)\in R} \sum_{i,a,z: x_i \neq y_i}
\sqrt{u(x,y,i)v(x,y,i)} (|\alpha_{i,a\oplus x_i,z}|
|\beta_{i,a\oplus y_i,z}|  + |\alpha_{i,a,z}| |\beta_{i,a,z}|)\\
= & \sum_{k=1}^T \sum_{i,a,z} \sum_{(x,y)\in R: x_i \neq y_i}
\sqrt{u(x,y,i)v(x,y,i)} (|\alpha_{i,a\oplus x_i,z}|
|\beta_{i,a\oplus y_i,z}| + |\alpha_{i,a,z}| |\beta_{i,a,z}|)\\
\end{array}
\]
by \eqref{eq:r1}. We then use inequality $2AB\leq A^2 + B^2$ to
get
\[
\sqrt{u(x,y,i)v(x,y,i)} |\alpha_{i,a\oplus x_i,z}|
|\beta_{i,a\oplus y_i,z}| \leq
\frac{1}{2}(u(x,y,i)\sqrt{\frac{v_{y,i}}{u_{x,i}}\frac{w_x}{w_y}}|\alpha_{i,a\oplus
x_i,z}|^2 + v(x,y,i)\sqrt{\frac{u_{x,i}}{v_{y,i}}\frac{w_y}{w_x}}
|\beta_{i,a\oplus y_i,z}|^2),
\]
\[
\sqrt{u(x,y,i)v(x,y,i)} |\alpha_{i,a,z}| |\beta_{i,a,z}| \leq
\frac{1}{2}(u(x,y,i)\sqrt{\frac{v_{y,i}}{u_{x,i}}\frac{w_x}{w_y}}|\alpha_{i,a,z}|^2
+ v(x,y,i)\sqrt{\frac{u_{x,i}}{v_{y,i}}\frac{w_y}{w_x}}
|\beta_{i,a,z}|^2),
\]
Denote $A = \min_{x,y,i: (x,y)\in R, x_i \neq y_i} \frac{w_x
w_y}{u_{x,i} v_{y,i}}$. Note that
\[
\sum_{y: (x,y)\in R, x_i \neq y_i} u(x,y,i) = u_{x,i},\ \sum_{x:
(x,y)\in R, x_i \neq y_i} v(x,y,i) = v_{y,i}
\]
by the definition of $u_{x,i}$ and $v_{y,i}$, we have
\[
\begin{array}{rl}
& (1-c) \sum_{(x,y)\in R}w(x,y) \\
\leq & \frac{1}{2} \sum_{k=1}^T \sum_{i,a,z} [\sum_{x\in X}
\sqrt{\frac{u_{x,i}v_{y,i}}{w_x w_y}} w_x (|\alpha_{i,a\oplus
x_i,z}|^2 + |\alpha_{i,a,z}|^2) \\
& \hspace{6em} + \sum_{y\in Y} \sqrt{\frac{u_{x,i}v_{y,i}}{w_x
w_y}} w_y (|\beta_{i,a\oplus y_i,z}|^2 + |\beta_{i,a,z}|^2)] \\
\leq & \frac{1}{2} \sum_{k=1}^T [\sum_{x\in X} \sqrt{1/A} w_x
\sum_{i,a,z} (|\alpha_{i,a\oplus x_i,z}|^2 + |\alpha_{i,a,z}|^2)
\\
& \hspace{4em} + \sum_{y\in Y} \sqrt{1/A} w_y \sum_{i,a,z}
(|\beta_{i,a\oplus y_i,z}|^2 + |\beta_{i,a,z}|^2)] \\
= & \sqrt{1/A} \sum_{k=1}^T (\sum_{x\in X} w_x + \sum_{y\in Y} w_y ) \\
= & 2 T \sqrt{1/A}\sum_{(x,y)\in R}w(x,y)
\end{array}\]
by noting that $\sum_x w_x = \sum_y w_y = \sum_{(x,y)\in
R}w(x,y)$. Therefore, $T = \Omega(\sqrt{A})$.
\end{proof}

We denote by $Alb_4(f)$ the best possible lower bound for function
$f$ achieved by this theorem. It is easy to see that $Alb_4$
generalizes $Alb_3$. $Alb_4$ may be easier to use than $Alb_3$.
However, according to Szegedy's recent result \cite{Sz04},
$Alb_3$, $Alb_4$ and the quantum adversary method proposed by
Barnum, Saks and Szegedy in \cite{BSS03} are all equivalent.

\vspace{1em} \noindent \textbf{Acknowledgement}

The author would like to thank Andrew Yao who introduced the
quantum algorithm and quantum query complexity area to me, and
made invaluable comments to this paper. Yaoyun Shi and Xiaoming
Sun also read a preliminary version of the present paper and both,
esp. Yaoyun Shi, gave many invaluable comments and corrections.
Thanks also to Andris Ambainis for telling that it is still open
whether $Alb_2(f)\leq \sqrt{C_0(f)C_1(f)}$ is true for total
functions, and to Mario Szegedy for sending me his note
\cite{Sz04}.


\begin{thebibliography}{}
\bibitem{YZ03} Andrew Yao, Shengyu Zhang, Graph property and
circular functions: how low can quantum query complexity go?
\end{thebibliography}
\end{document}